\documentclass[fleqn,usenatbib,usedcolumn]{mnras}
\usepackage{aas_macros}
\usepackage{cleveref}
\usepackage{graphicx}	   
\usepackage{amsmath}	   
\usepackage{amssymb}	   
\usepackage{multicol}      
\usepackage{color}
\usepackage[utf8]{inputenc}
\usepackage[english]{babel}
\usepackage{bm}		       
\usepackage{listings}
\usepackage[skins,theorems]{tcolorbox}
\usepackage{array}
\usepackage{booktabs}

\let\en=\ensuremath
\newcommand{\ve}[2]{\en{#1_1},~\en{#1_2},\ \ldots,~\en{#1_{#2}}}
\newcommand{\ma}[3]{\en{#1_{{#2}1}},~\en{#1_{{#2}2},\ \ldots,~\en{#1_{{#2}#3}}}}

\usepackage[T1]{fontenc}
\usepackage{ae,aecompl}
 
\usepackage{mathptmx}
\usepackage{txfonts}

\title[]{Is the cluster environment  quenching the  Seyfert  activity  in elliptical and spiral galaxies?} 

\author[R.~S.~de~Souza et al.]{R.~S.~de~Souza$^{1}$\thanks{E-mail:rafael@caesar.elte.hu}, M.~L.~L.~Dantas$^{2}$,  A.~Krone-Martins$^{3}$, E.~Cameron$^{4}$, P.~Coelho$^{2}$,  
\newauthor 
M.~W.~Hattab$^{5}$,  M.~de~Val-Borro$^6$,
J.~M.~Hilbe$^{7}$, J.~Elliott$^{8}$ and A.~Hagen$^{9}$,
\newauthor
for the COIN Collaboration\\
$^{1}$MTA E\"otv\"os University, EIRSA ``Lendulet'' Astrophysics Research Group, Budapest 1117, Hungary\\
$^{2}$Instituto de Astronomia, Geof\'isica e Ci\^encias Atmosf\'ericas, Universidade de S\~ao Paulo, 
R. do Mat\~ao 1226, 05508-090, S\~ao Paulo, Brazil\\
$^{3}$CENTRA/SIM, Faculdade de Ci\^encias, Universidade de Lisboa, Ed. C8, Campo Grande, 1749-016, Lisboa, Portugal\\
$^{4}$Department of Zoology, University of Oxford, Tinbergen Building, South Parks Road, Oxford, OX1 3PS, United Kingdom\\
$^{5}$Center for Biomarker Research and Personalized Medicine, Virginia Commonwealth University, Richmond, VA, USA\\
$^{6}$Department of Astrophysical Sciences, Princeton University, Princeton, NJ 08544, USA\\
$^{7}$Department of Statistics, School of SFD, Arizona State University, Tempe, AZ, USA\\
$^{8}$Harvard-Smithsonian Center for Astrophysics, 60 Garden St., Cambridge, MA 02138, USA\\
$^{9}$Department of Astronomy and Astrophysics, Pennsylvania State University, University Park, PA 16802, USA
}

\date{Last updated \today; in original form 2 \today}

\pubyear{2016}

\begin{document}
\label{firstpage}
\pagerange{\pageref{firstpage}--\pageref{lastpage}}
\maketitle

\begin{abstract}
We developed a hierarchical Bayesian model (HBM) to investigate how the presence of Seyfert activity relates to  their environment, herein represented by the galaxy cluster mass, $M_{200}$, and the normalized cluster centric distance, $r/r_{200}$. We achieved this by constructing an unbiased sample of galaxies from the \textit{Sloan Digital Sky Survey}, with morphological classifications provided by the \textit{Galaxy Zoo Project}. A propensity score matching  approach is introduced to control  the effects of confounding variables: stellar mass, galaxy colour, and star formation rate. The connection between Seyfert-activity and environmental properties in the de-biased sample is modelled within an HBM framework using the so-called  logistic regression technique, suitable  for the analysis of binary data (e.g., whether or not a galaxy hosts an AGN).
Unlike standard ordinary least square fitting methods, our methodology naturally allows modelling the probability of Seyfert-AGN activity in galaxies  on their natural scale, i.e.\ as a binary variable. Furthermore, we demonstrate how an HBM can incorporate information of each particular galaxy morphological type in an unified framework.
In elliptical galaxies our analysis indicates a strong correlation of Seyfert-AGN activity with $r/r_{200}$, and a weaker correlation with the mass of the host cluster. 
In spiral galaxies these trends do not appear, suggesting that the link between Seyfert activity and the properties of spiral galaxies are independent of the environment.
\end{abstract}



\section{Introduction}

For a long time it has been argued that active galactic nuclei (AGN) are powered by the accretion of gas into a super-massive black hole (SMBH) located at the centre of its host galaxy \citep[e.g.][]{Lynden-Bell+1969,Magorrian1998,OrbandeXivry2011}. The AGN-driven energetic outflows, that are launched from the accretion disc surrounding the SMBH,  are fundamental processes thought to shape the formation of massive galaxies. In particular, AGN feedback is often invoked in semi-analytic models to explain the suppression of gas accretion on to the host galaxy and the subsequent quenching of star formation \citep{Birnboim.Dekel:2003,Keres.etal:2005,Dekel.Birnboim:2006,Somerville2008,Cattaneo.etal:2009,Keres.etal:2009,Cattaneo.etal:2011}. Empirical evidence supporting the idea that nuclear activity and the evolution of the host galaxy are closely intertwined includes the tight correlation between SMBH mass and the  bulge velocity dispersion.   The latter  is taken to imply a strong connection between the formation and evolution of the central SMBH and that of the bulge \citep[e.g.,][]{FerrareseMerritt2000,Gebhardtetal2000,Tremaine2002,HaringRix2004,Somerville2008,Reines2015}. The AGN feedback interacts with the gas of its host via radiation pressure, winds and jets, hence helping to shape the final mass of the stellar components \citep{Fabian2012}. The AGN feedback can also affect other galaxies in the surrounding environment \citep{Ishibashi2016}, with deep implications in the galaxies, groups and cluster evolution.
However, a detailed picture of how exactly AGN can affect the evolution of its host remains to be established.

The physical mechanisms that power the AGN are also a matter of debate. Some studies indicate that the dissipation of angular momentum during major mergers might allow gas to accrete into the central black hole \citep[e.g.,][]{Hopkins.etal:2006, Kaviraj.etal:2015}. Other secular mechanisms, such as disc/bar instabilities, colliding clouds, and supernovae explosions have also been proposed as AGN activity triggers \citep[see][for reviews]{Kormendy.Kennicutt:2004, Martini:2004, Jogee:2006,Booth2013}. Therefore, one should expect to find a correlation between AGN activity and the visual morphology of its host galaxy.
This connection has been explored in several studies. For instance, \citet{OrbandeXivry2011} found that the hosts of Narrow-line Seyfert 1 (NLS1)  systems tend to be very late-type galaxies, such as grand design spirals, and that  
NLS1 may  represent a different class of AGN in which the black hole
growth is dominated by secular evolution much more than their broad-Line Seyfert 1s counterparts.
 \citet{Villarroel2014} found differences in the colour distribution and AGN activity of the neighbours to Type-1 and Type-2 AGN and in the fraction of AGN residing in spiral hosts depending on  presence or not of a neighbour. 
It is worth noting that studies as the above implying that  different AGN types do not interact in the same way with their environment represent  potential issues to the so-called unification AGN model \citep{Antonucci1993,Urry1995}, which suggests that all AGN are the same type of object viewed from a different angle. 
 \citet{Alonso.etal:2014} show that spiral galaxies hosting AGN in groups are more likely to be barred than their counterparts in the field. Similar results were presented by \citet{Coelho.Gadotti:2011}, who found twice as many AGN among barred galaxies, as compared to their unbarred counterparts, for low mass bulges. Nevertheless, \citet{Cisternas.etal:2015} claim that the presence of a bar has no influence on the AGN strength, with barred and unbarred active galaxies showing equivalent X-ray luminosity distributions, and  \cite{Miller.etal:2003} also found that the AGN fraction is independent of the  morphological type of the host galaxy, indicating that there is no overall relationship between the star formation activity in the disc component of galaxies and the presence of an AGN. 
However, \citet{wada04} found evidence that mass accretion is not constant during an activity cycle of $10^8$ years, but is composed of several shorter episodes. This difference in time-scales may explain the lack of success in finding a correlation between the presence of structures at kpc scales (for example, disc/bar interactions) with the emerging activity in galaxies, except for  bright objects such as Quasars \citep[e.g.,][]{moles+95,krongold+01}. 
Finally, \citet{Schawinski2010} found that in early-type galaxies, it is preferentially those with the least massive central black holes that are active, unlike late-type galaxies, in which the most massive black holes are the active ones. Therefore, it is still unclear to what strength the AGN activity is connected to the presence of particular morphological features. 

Environmental effects could also turn on or off the AGN activity. Instabilities originating from galaxy mergers, and from interactions between the galaxy and the cluster potential could drive gas towards the galaxy centre, powering the AGN. Notwithstanding, the gas reservoirs of galaxies may be stripped by the tidal field of the group/cluster environment \citep[e.g.]{Larson.etal:1980,Roediger2005,Roediger2015}, by the numerous high-speed encounters with smaller galaxies \citep[e.g.,][]{Moore.etal:1996}, or by the ram pressure of the intra-cluster gas \citep[][]{Gunn.Gott:1972,Roediger2005,Roediger2015}. The absence of gas to fuel the central black hole would inevitably lead to the turn off of the nuclear activity. \citet{vonderLinden.etal:2010} found that the fraction of star-forming galaxies hosting a powerful optical AGN is independent of the normalized cluster-centric distance,  $r/r_{200}$,\footnote{The quantity $r_{200}$ is  the radius  inside which the mean density is 200 times the critical density of the Universe at the cluster redshift.} indicating that the link between star formation and AGN in these galaxies does not depend on the environment. Nonetheless, the fraction of red galaxies which host a weak optical AGN decreases towards the cluster centre, following the trend of star-forming galaxies. This might indicate that environmental effects gradually quench both the star formation and the AGN activity.
More recently \citet{Pimbblet2013} using sample of cluster relatively free from mergers, which can  locally
enhance AGN activity, found a strong relation between AGN activity and $r/r_{200}$, with significant increase of AGN fraction from the cluster centre to 1.5 Virial radii, with  massive galaxies  systematically hosting a larger fraction of AGN at any radial location.  

Moreover, it is well known that many galaxy properties, such as stellar mass \citep[$M_{*}$;][]{Schawinski2010}, morphology \citep{Dressler:1980,Calvi2012}, star formation rate \citep[SFR;][]{Abraham.etal:1996,Hashimoto.etal:1998,
Gomez.etal:2003,Kauffmann.etal:2004,Harris2016}, and optical colours \citep{Strateva2001,Hogg.etal:2004,Kauffmann.etal:2004,Blanton.etal:2005,Baldry.etal:2006} are strongly correlated with the environment where the galaxy resides. Therefore, the interplay between these competing processes results in a very intricate relation between the AGN activity, galaxy properties and local environment. This requires a careful statistical modelling and the construction of an unbiased sample in order to make robust statistical and, consequently, physical claims. 

In this work, we developed a hierarchical Bayesian model (HBM) to explore the  roles of  morphology, environment, and the occurrence of Seyfert galaxies, $f_{\rm Seyfert}$. We use a logistic regression methodology, that is a technique designed to deal with binary data \citep[e.g.][]{deSouza2015}. Additionally, to ease the biases caused by competing effects, we applied a propensity score matching technique (PSM) to build a control sample of inactive galaxies with similar colours $(g-r)$, $M_{*}$, and SFR.

The outline of this paper is as follows. In Section~\ref{sec:data} we provide an overview of the  sample selection,  the derivation of galaxy and cluster properties, and the selection of the AGN sample via emission line diagnostic diagrams.  Section~\ref{sec:method} gives a description of the Bayesian statistical methodology.  We present our results in depth in Section~\ref{sec:res}, and, finally, we discuss their physical motivations along with our conclusions in Section~\ref{sec:results_discussion}.

\section{Cluster sample}
\label{sec:data}

In this section we describe our dataset, which was selected from the updated
version of the catalogue compiled by \citet{Yang.etal:2007}(
communication). The sample comprises groups and clusters of
galaxies from the \textit{Sloan Digital Sky Survey} 7\textsuperscript{th} 
\citep[SDSS-DR7,][]{Abazajian.etal:2009}, with additional data  retrieved from
the SDSS 12\textsuperscript{th} data release database
\citep[SDSS-DR12,][]{Alam.etal:2015}. We selected all halos within a redshift
range of $0.015 < z < 0.1$, having  at least 10 galaxy members, and with $13.4
< \log M_{200} < 14.6$, in which  $M_{200}$ corresponds to the mass within
spheres that are 200 times denser than the critical density of the Universe at
the cluster redshift\footnote{All reported masses  are in  Solar mass units,  but suppressed from the text for simplicity.}. 
We included galaxies up to $\sim 10\cdot r_{200}$ by applying an assignment scheme similar to the one described in \citet{Yang.etal:2007} and \citet{Duarte.Mamon:2015}. 
As for the $M_{*}$ and SFR, we adopted the  estimated values from \citet{Brinchmann.etal:2004}.  
Our sample includes only galaxies brighter than ${M_{p,r}<-20.4}$, in which ${M_{p,r}}$ is the $k$-corrected SDSS Petrosian absolute magnitude in the $r$-band. This magnitude corresponds to the 95 per cent completeness limit of the sample. The $k$-corrections were obtained with the {\tt kcorrect} code (version 4\_2) described by \citet{Blanton.Roweis:2007}, choosing as reference the median redshift of the SDSS main galaxy sample ($z=0.1$).
All the criteria described above leads to a sample of  $32,353$ galaxies within  $1,122$ groups and clusters. 

We also split the sample into two groups, ellipticals and spirals, in order to avoid the oversimplified characterization of the AGN host galaxies as a single class, but rather understand how the galaxy-AGN co-evolution varies between morphologies \citep{Schawinski2010}. This also helps to account for the well known  morphology-$r/r_{200}$ relation, in which spirals tend to be mainly located in the outskirts of their respective groups or clusters, whereas ellipticals can be found specially in their inner parts towards regions of increasing local galaxy density \citep[e.g.][]{Goto2003}.  The morphological typing relies in the visual classification scheme performed by citizen scientists from the \textit{Galaxy Zoo Project} \citep{Lintott.etal:2008}. We use the definition of clean sample of \citet{Land2008}, which requires at least 80 per cent majority agreement on the morphology of any object. For the analysis,  we place all clusters into a common scale  (i.e. $r_{200}$) to form a composite sample in order to  minimize radial sampling bias due to variations in cluster richness \citep[e.g.][and references therein]{Barkhouse2009}.
\footnote{A detailed description of the construction of the entire dataset is  available  in \citet{Trevisan2016}.}

\subsection{AGN selection}

\begin{figure*}
\resizebox{1\hsize}{!}{\includegraphics{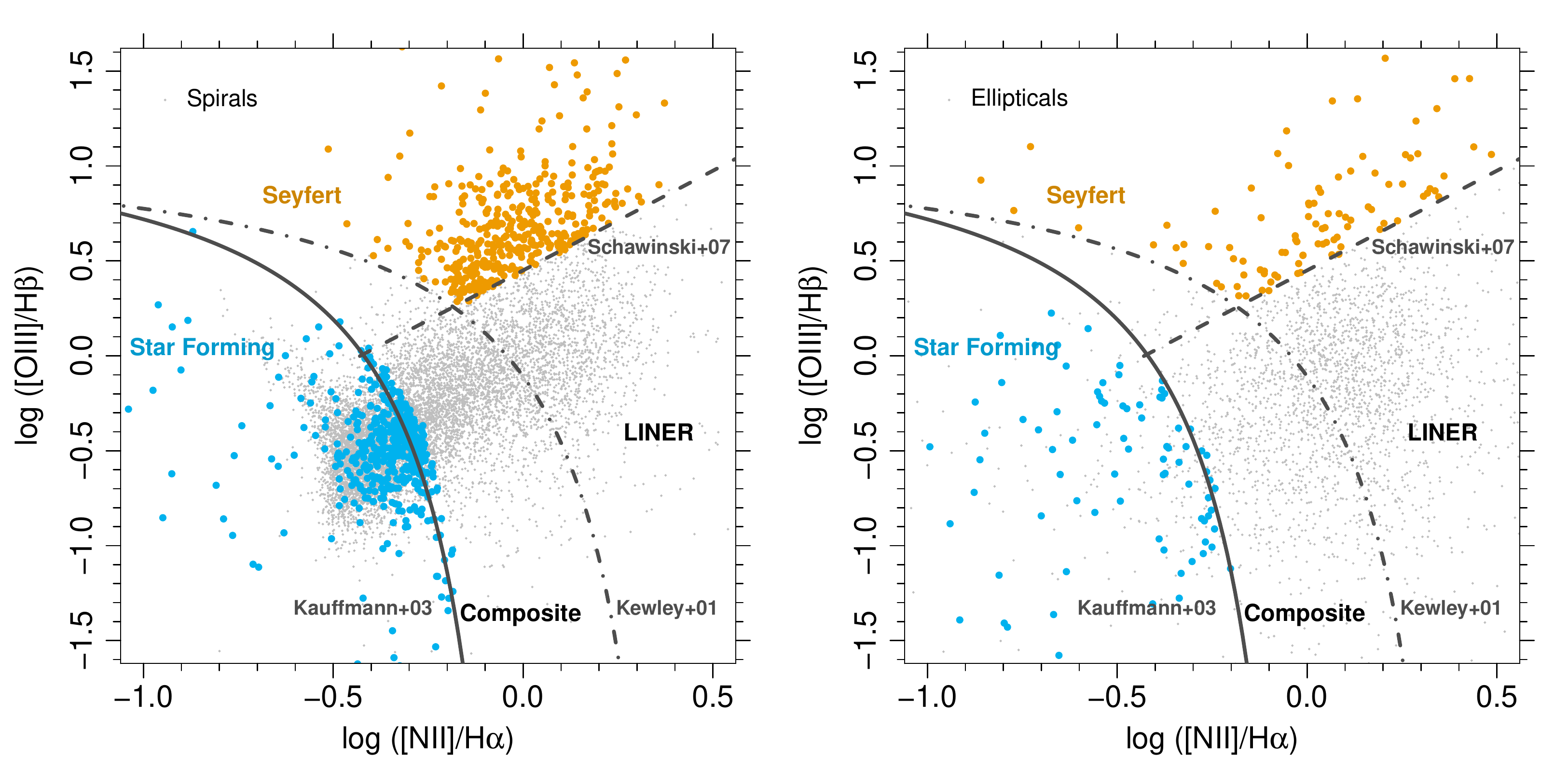}} 
\caption{BPT plane for our  galaxy sample for both spirals (left panel) and elliptical galaxies (right panel). The vertical-axis represent the ratio of $[\mathrm{OIII}]/H_{\beta}$, while the horizontal-axis is the ratio of $[\mathrm{NII}]/H_{\alpha}$. The grey dots represent the entire galaxy sample, the orange dots are the Seyfert galaxies, and the blue points represent the control sample of inactive galaxies selected with the propensity score matching (see Sec.~\ref{sec:psm}). The solid curve is the \citet{Kauffmann.etal:2003} line: galaxies above the curve are designated AGN, those below are regular star-forming galaxies. The dot-dashed line represents the \citet{Kewley.etal:2001} curve, where galaxies between the \citeauthor{Kauffmann.etal:2003} and \citeauthor{Kewley.etal:2001} curves are defined as composites: weaker AGN whose hosts are also star-forming galaxies. The dashed line is the \citet{Schawinski.etal:2007} curve, which separates LINER and Seyfert objects.  
}
\label{fig:BPT}
\end{figure*}

The classification of  objects in our sample is based on the diagram introduced by \citet{Baldwin.etal:1981}, hereafter BPT [but see also \citet{VeilleuxOsterbrock1987,Kewley.etal:2001,Kauffmann.etal:2003,Stasinska+06,Schawinski.etal:2007}], as shown in Figure~\ref{fig:BPT}, and is available for galaxies with the emission lines H$\beta$, [OIII], H$\alpha$, [NII] for which the signal to noise ratio $S/N > 1.5$.   The Star Forming galaxies are those whose  emission lines are partially or fully dominated by star formation in the diagram as defined by the theoretical extreme starburst line of \citealt{Kewley.etal:2001}. Composite objects are the ones falling between this extreme starburst line and the empirical pure starburst line of \citet{Kauffmann.etal:2003}. These are objects that are constituted by coexisting/competing star formation and nuclear activity in terms of ionizing luminosity strength. On the right side of the \citet{Kewley.etal:2001} line reside the galaxies whose emission lines are dominated by sources of ionization other than young stars, which are empirically divided into two major AGN sub-classes, the lower branch of  low-ionization narrow emission-line regions (or LINERs) and the upper branch of Seyfert galaxies delimited by the line derived by \cite{Schawinski.etal:2007}. This leads to a classification of objects in our sample as Star Forming, Seyfert, LINER,  and Composite as displayed in Fig.~\ref{fig:BPT}, where we show the positions of all galaxies in the sample on the BPT diagram.

Nonetheless, the nature of LINERs, as low-luminosity AGN, is still uncertain and there is a growth of evidence indicating that the majority of galaxies with such spectra classification is unlikely to be true AGN \citep[see e.g.,][]{Stasinska+2008,Schawinski2010,Singh.etal:2013}. Many of them may actually be retired galaxies powered by old stellar populations \citep{Maraston2005,CidFernandes2010,Stasinska2015}. 
Due to this uncertain nature of LINER galaxies, and with the intent to reduce contamination from false-negative AGN in objects belonging to the Composite region, we chose to eliminate most of them from our analysis, but allowing a narrow region around the \citet{Kauffmann.etal:2003} curve ($\Delta \log([NII]/H\alpha) \lesssim 0.05$) to accommodate inherent uncertainties in the curve definition. This allows us to retain  only the clearly distinct AGN and non-AGN host galaxies in this study, following an  approach similar to the one employed by \citet{Schawinski2010}.

\subsection{Control sample}
\label{sec:psm}

\begin{figure*}
\resizebox{1\hsize}{!}{\includegraphics[trim={0.1cm 0.1cm 0.1cm 0.1cm},clip]{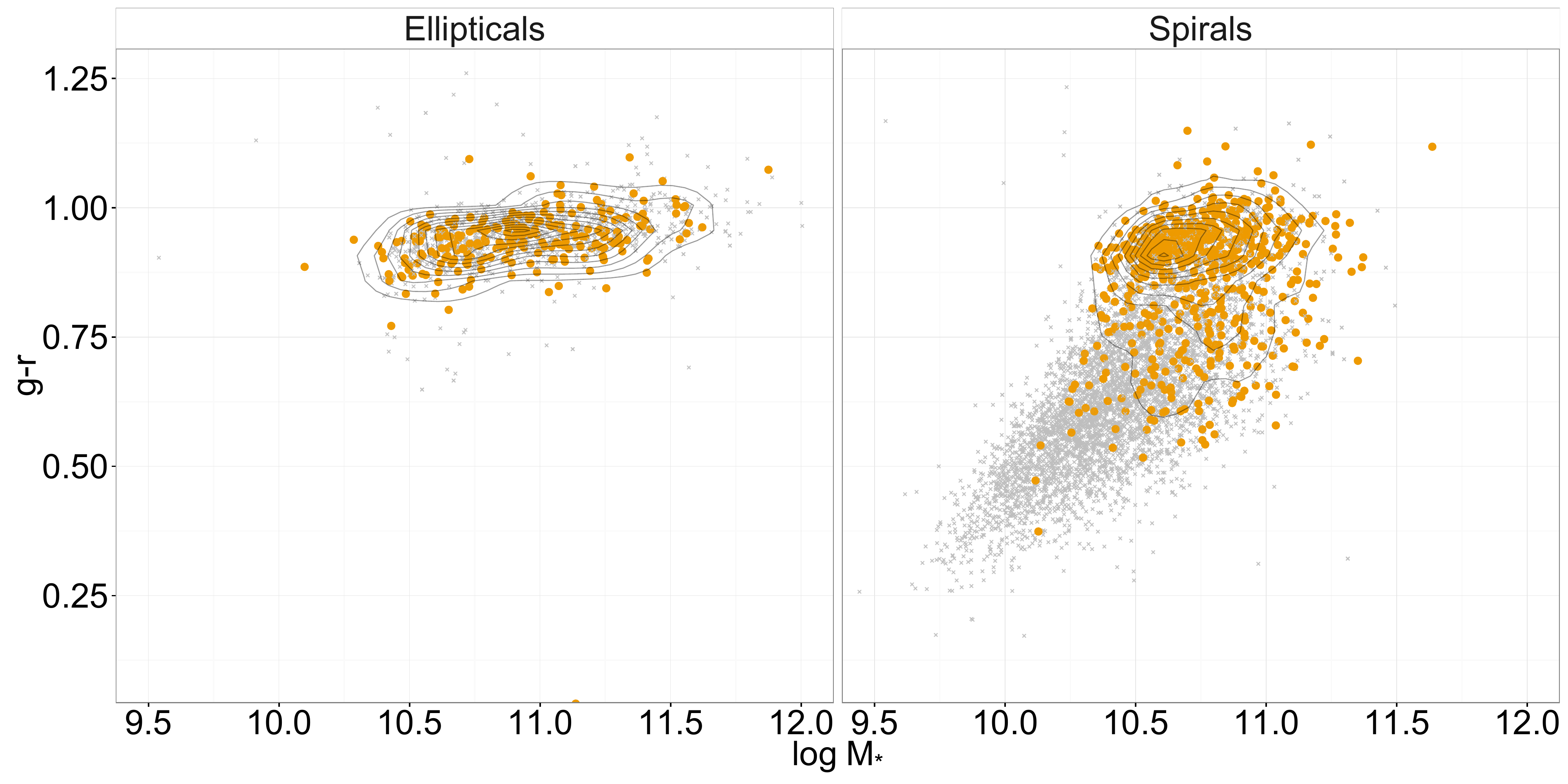}}
\resizebox{1\hsize}{!}{\includegraphics[trim={0.1cm 0.1cm 0.1cm 0.1cm},clip]{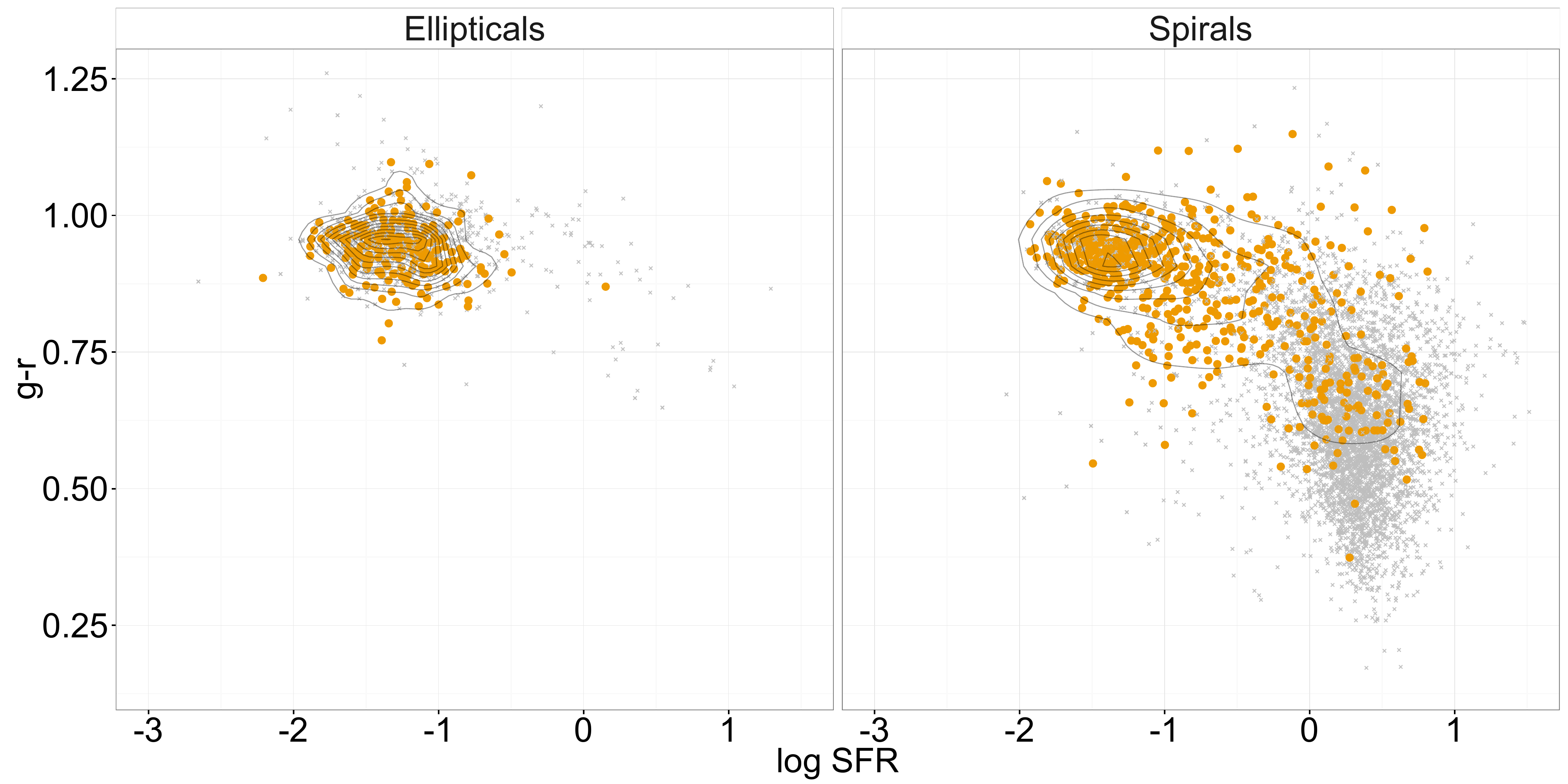}}
\caption{Distribution of the Seyfert-AGN (orange dots) on the $(g-r)$ colour-mass (top panel), and colour-star-formation (bottom panel) diagrams. Left panels: elliptical galaxies; right panels: spiral galaxies. The grey dots represent the whole galaxy population, on top of which we plot the Seyfert hosts, with companying grey contour levels of the Seyfert-population.
}
\label{fig:colour_mass}
\end{figure*}

Galaxies hosting AGN have different characteristics from their inactive counterparts, preferentially  populating the so-called \emph{green valley} and \emph{red sequence} of the colour-mass diagram \citep[e.g.][]{Smolvic2009,Schawinski2010}. Additionally, in recent years, there has been an ongoing discussion regarding AGN feedback and the connection with SFR: it is believed that AGN activity suppresses the formation of new stars~\citep{Page2012,Booth2013,Bluck2014,Li2015}. Given that spirals are usually rich in gas and dust, and consequently have high levels of SFR, it is possible to observe a peak of SFR in spirals that do not host AGN. Moreover,  the distribution of galaxy colours is linked to the SFR distribution: bluer galaxies possess younger stellar populations, whereas spirals which host AGN are redder and have older stellar populations.  These trends  can be seen  in our sample, both in the colour-mass and colour-SFR diagram  in the top and bottom panels of Fig.~\ref{fig:colour_mass}, respectively. The galaxies hosting Seyfert-AGN (orange dots) are compared against inactive galaxies in the sample represented  by the bulk of grey dots. 

\begin{figure*}
\resizebox{0.425\hsize}{!}{\includegraphics[trim={0.1cm 1.825cm 1.25cm 2cm},clip]{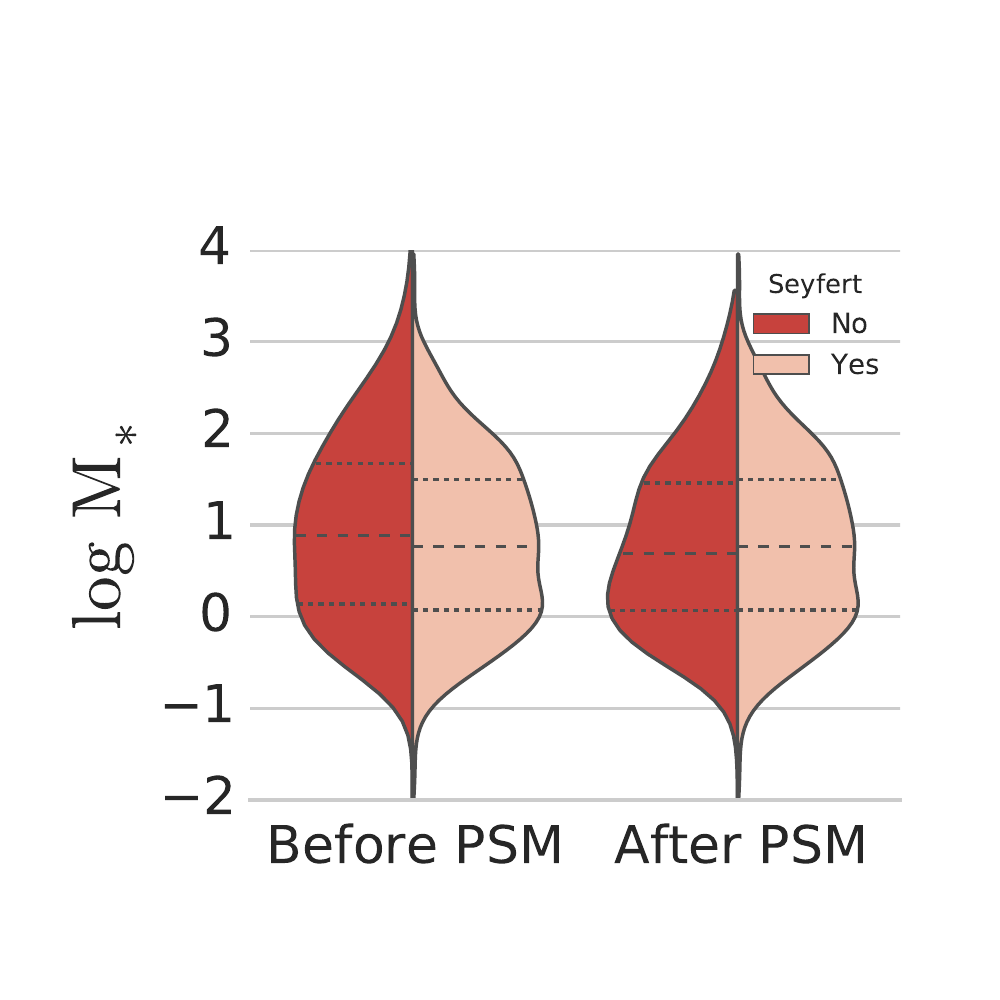}}
\resizebox{0.425\hsize}{!}{\includegraphics[trim={0.1cm 1.825cm 1.25cm 2cm},clip]{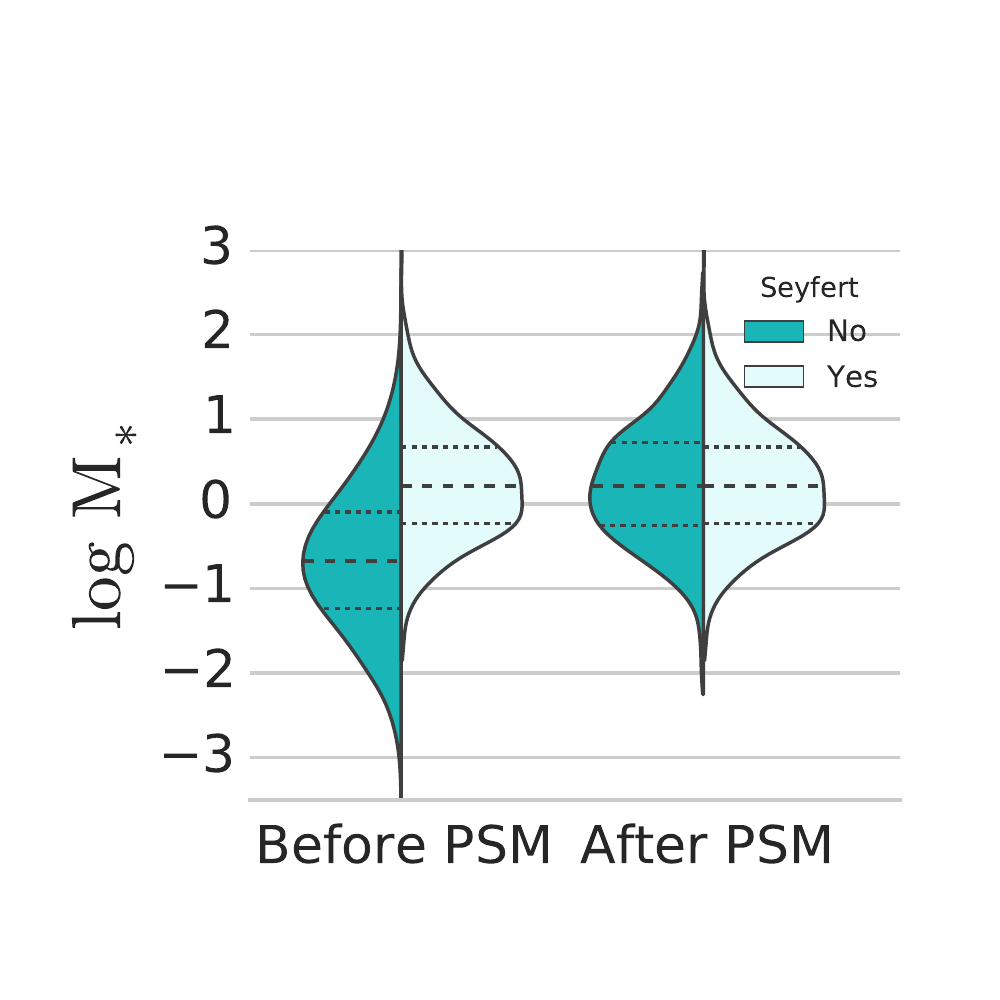}}
\resizebox{0.425\hsize}{!}{\includegraphics[trim={0.1cm 1.825cm 1.25cm 2cm},clip]{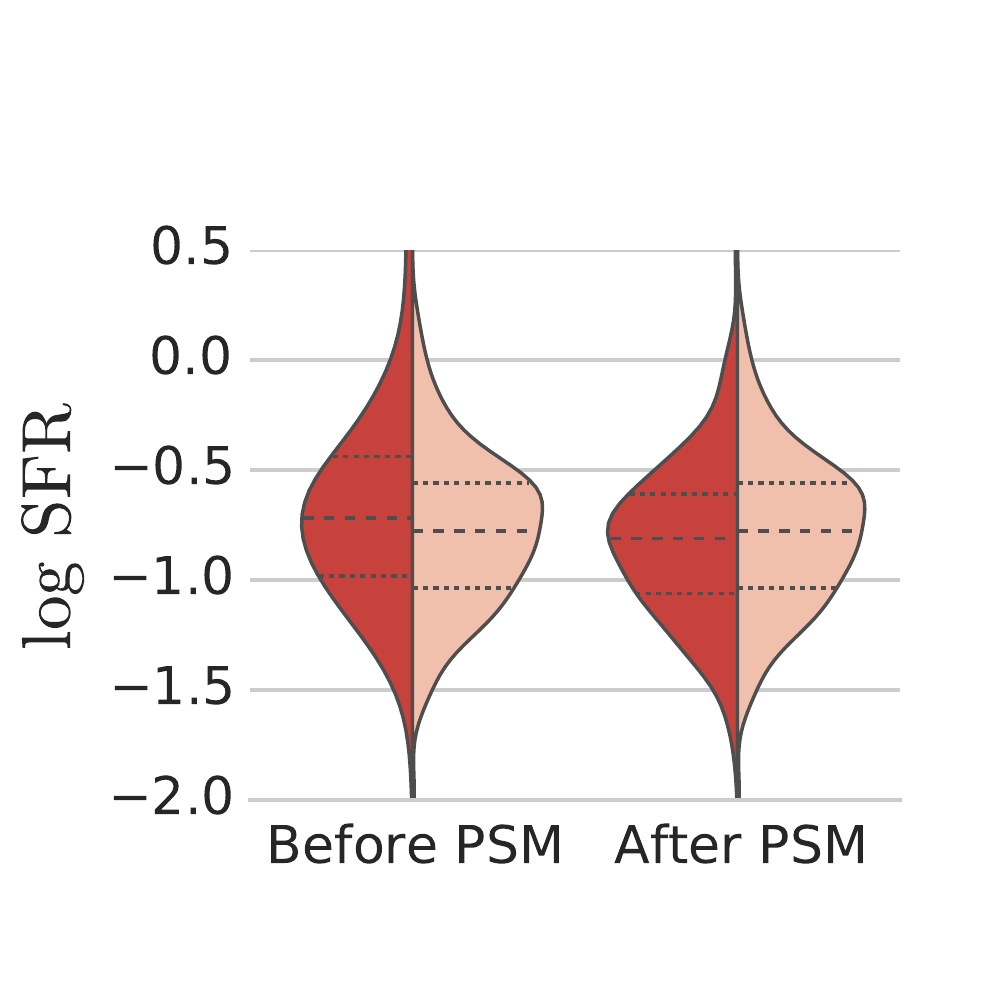}} 
\resizebox{0.425\hsize}{!}{\includegraphics[trim={0.1cm 1.825cm 1.25cm 2cm},clip]{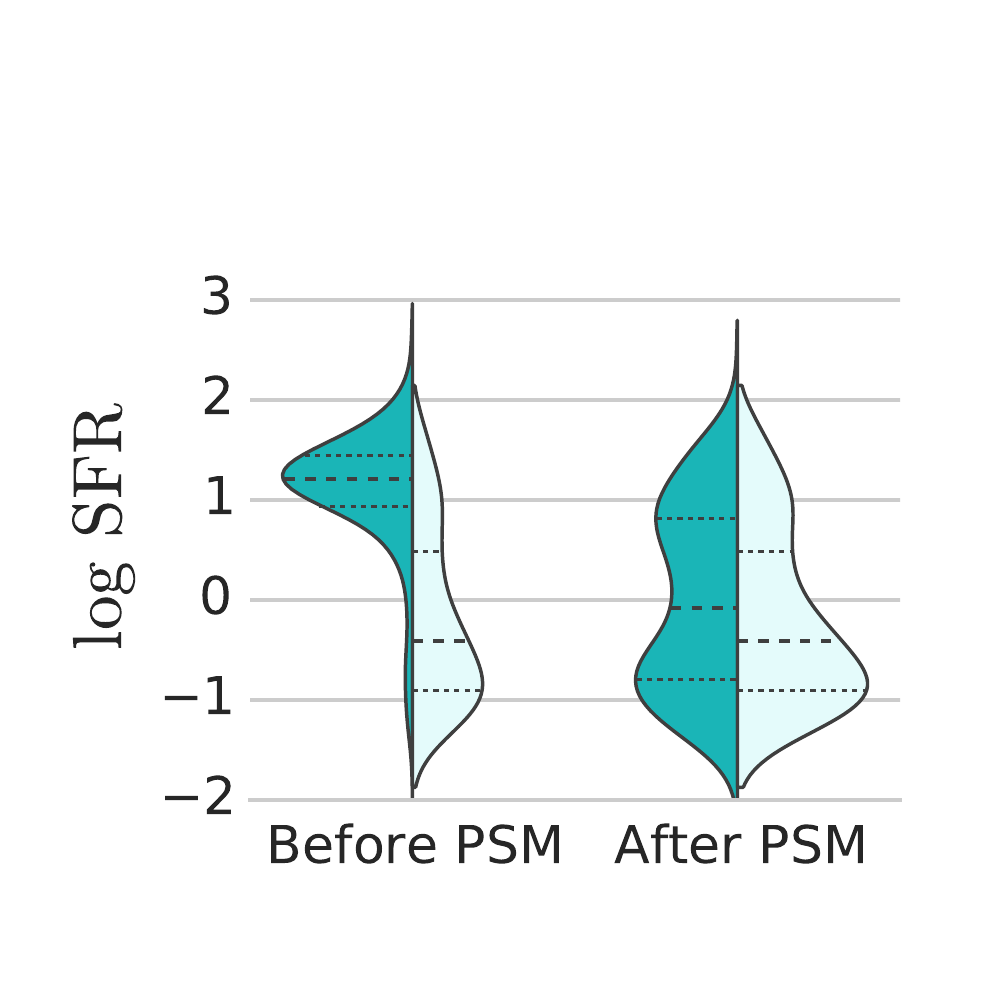}} 
\resizebox{0.425\hsize}{!}{\includegraphics[trim={0.1cm 1cm 1.25cm 2cm},clip]{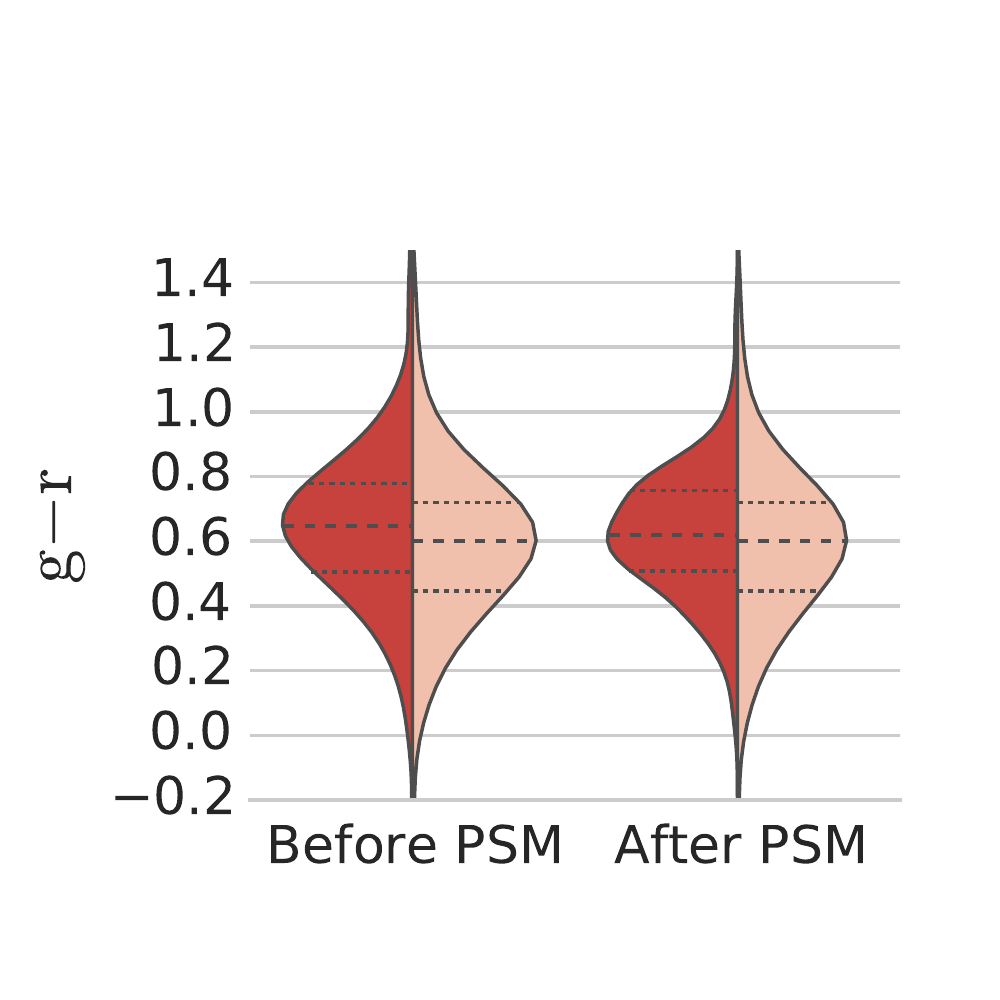}}
\resizebox{0.425\hsize}{!}{\includegraphics[trim={0.1cm 1cm 1.25cm 2cm},clip]{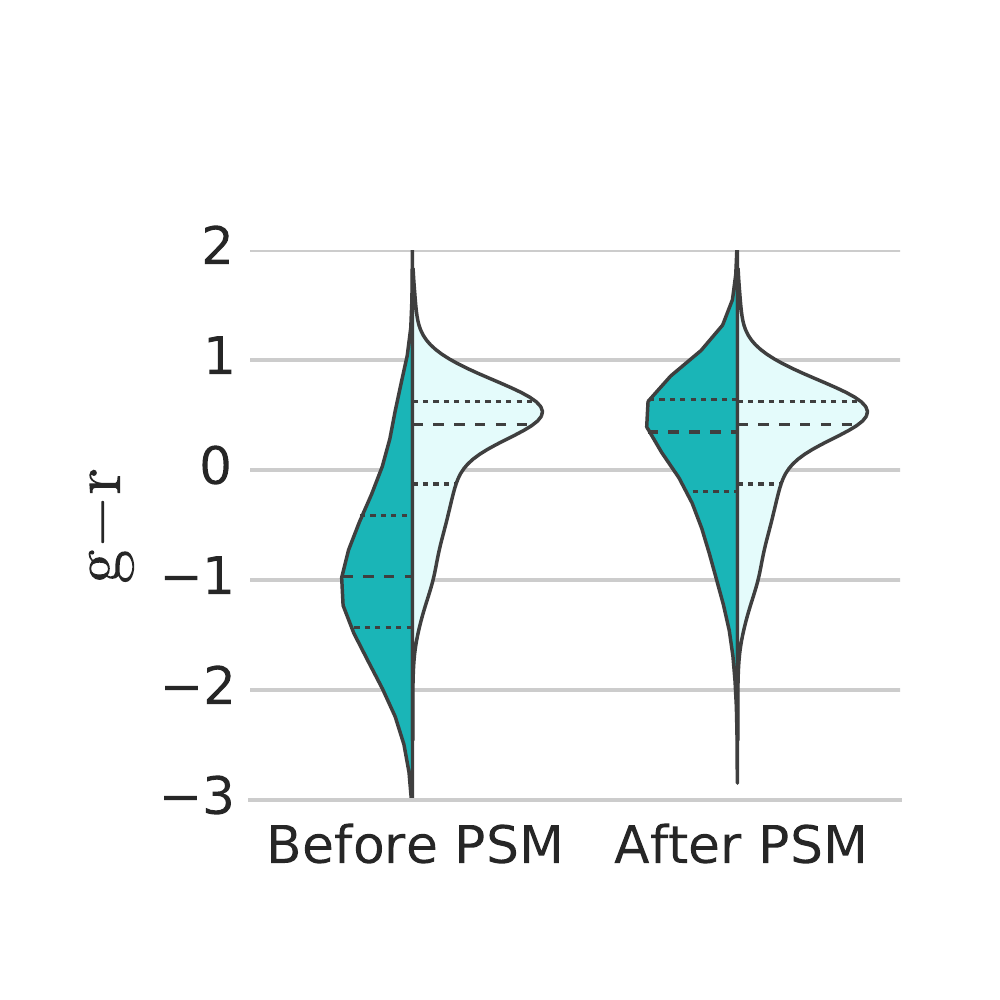}}
\caption{Distribution of $\log \, M_{*}, \log$ SFR, and the $(g-r)$ colour for elliptical (left column, in red) and spiral (right column, in cyan) galaxies. In each representation, the right and left distributions represent respectively the samples with and without Seyfert AGN activity, portrayed by an asymmetric violin plot. The dotted and solid lines in each violin represent the median and 95 per cent quantiles of the sample, while the shape, denotes its  probability distribution. Note that all variables are standardized, for illustrative purposes, as follows, $x^* = (x-\hat\mu_x)/\hat \sigma_x$, for $x^*$ is the standardized variable (vertical axis), and $\hat \mu_x$ and $\hat \sigma_x$ represent its sample mean and standard deviation respectively.}
\label{fig:PSM}
\end{figure*}

To mitigate this bias, mostly due to the spirals, we built a control sample of
inactive galaxies (i.e.  those under the curve described by
\citealp{Kauffmann.etal:2003}) by matching each pair Seyfert/non-Seyfert
galaxy  against the confounding covariates, $X_{c} = \left\lbrace M_{*}, (g-r),
SFR\right\rbrace$, and their morphological type\footnote{Although $(u-r)$ is
the most commonly used SDSS colour to characterize galaxies through the
colour-mass diagram \citep[e.g.][]{Strateva2001,Baldry2004,Baldry.etal:2006},
we chose to use the optical colour $(g-r)$ due to the ageing effects and
processes suffered by the $u$ band, which can be non-negligible for SDSS-DR7
data \citep[see e.g.][]{Doi2010}.}. The match is performed via a non-parametric
approach known as propensity score matching
\citep[PSM,][]{HoImaKin07,Austin2011} using the {\sc matchit} {\sc R} package
\citep{Ho2011}. For each Seyfert galaxy, the method searches for the closest
inactive galaxy,  in the multidimensional space formed by $X_{c}$, via a
$k$-nearest neighbourhood algorithm \citep[see  e.g. \S 3.1;][for a
review]{Ishida2013}; the results are presented in Fig.~\ref{fig:PSM}. After the
PSM, the distributions of the active and inactive sample closely resemble each
other and any potential  effect that $X_{c}$ may have on the presence of a
Seyfert-AGN is subsumed, since $X_{c}$ is virtually held constant within each
pair of galaxies. The final de-biased sample is composed by 1,744 objects from
which 492 are ellipticals and 1,252 are spirals. The subset is displayed in the
BPT diagram (Fig.~\ref{fig:BPT}) as follows: cyan dots represent the control
sample; the orange ones, the active sample. In the following section, we
describe how to build a Bayesian model to explore the effects of $r/r_{200}$
and $M_{200}$ in our de-biased sample.

\section{Bayesian  Logistic  Model}
\label{sec:method}

The simple \textit{linear regression} model has long been a mainstay of astronomical data analysis \citep[e.g.][]{Isobe1990,Feigelson1992,Kelly2007,Sereno2016}. One important example is the determination of the line of best fit through Hubble's diagram \citep[][more recently]{hub29,Abraham2001}. Nonetheless, the assumptions of the linear model fall short when the data to be modelled come from \textit{exponential family} distributions other than the Normal/Gaussian\footnote{The exponential family comprises a diverse set of common distributions describing both continuous and discrete random variables (e.g., Gaussian, Poisson, Bernoulli, Gamma, etc.)} \citep{Hil12,Hilbe2014}. 
For such  problems, there is an overarching solution  known as generalized linear models (GLMs). Regression models in the class of GLMs, first developed by \citet{nel72}, take a more general form than in ordinary linear regression:
\begin{align}
\label{eq:glm}
&Y_i \sim f(\mu_i, a(\phi) V(\mu_i)) , \notag \\
&g(\mu_i) = \eta_i,  \\ 
&\eta_i \equiv \boldsymbol{x}_i^T \boldsymbol{\beta} = \beta_0+\beta_1x_1+\cdots+\beta_kx_k. \notag 
\end{align}
In equation~(\ref{eq:glm}), $f$ denotes a response variable distribution from the exponential family (EF), $\mu_i$ is the response variable mean, $\phi$ is the EF dispersion parameter in the dispersion function $a(\cdot)$, $V(\mu_i)$ is the response variable variance function, $\eta_i$ is the linear predictor,  
$\boldsymbol{x}_i^T=\{\ma x i k \}^T$
is a vector of explanatory variables (covariates or predictors), 
$\boldsymbol{\beta}=\{\ve \beta k \}$
is a vector of covariate coefficients, and $g(\cdot)$ is the link function, which connects the mean to the linear predictor $\eta$. 
If the response is Gaussian, then $f$ = Normal,  $g(\mu) = \mu$, $a(\phi) =1$, $V(\mu)= \sigma^2$,   and   the general form of the GLM recovers the ordinary  linear regression as a special case:
\begin{align}
\label{eq:lm}
&Y_i \sim \mathrm{Normal}(\mu_i, \sigma^2) , \notag \\
 &\mu_i =  \beta_0+\beta_1x_1+\cdots+\beta_kx_k.  
\end{align}
To date, there has been only a handful of astronomical studies applying GLM techniques, such as logistic  \citep[e.g.,][]{Rai12,Rai14,deSouza2014}, Poisson \citep[e.g.,][]{and10}, gamma  \citep{Elliott2015}, negative binomial   \citep{deSouza2015}, and  Normal regressions \citep{Bhardwaj2016,Lenz2016}.   
The methodology discussed herein focuses on logistic regression, which is  suitable for handling Bernoulli (or binomial) distributed data. The Bernoulli distribution describes a process in which there are only two possible outcomes: success or failure (yes/no, on/off, red/blue, etc.; typically coded as 1/0).  
Bernoulli distribution is a particular case of the more general binomial
distribution, $\mathrm{Binomial}(n,p) =  \binom {n} {y} p^y (1-p)^{n-y}$, for
which $y$ is the number of successes ($y = 1$),   $n$ is the number of trials, and $p$ is the probability of success.  For the Bernoulli
distribution, $\mathrm{Bernoulli}(p) =  p^y (1-p)^{1-y}$, the number of trials, $n$,
is set to 1. The link function (Eq.~\ref{eq:logit}) derives directly from the
underlying Bernoulli probability distribution and ensures a bijection between
the $(-\infty,\infty)$ range of the linear predictor, $\eta$, and the (0,1)
range of non-trivial probabilities for the binomial population proportion (the
Bernoulli $p$). The natural link function for the Bernoulli distribution is
known as the logit link, which defines the logistic model:
\begin{equation}
\label{eq:logit}
g(p)= \mathrm{logit}(p) \equiv \log \left(\frac{p}{1-p}\right).
\end{equation}
The logit function has another desirable property: that the exponentiated coefficients of the linear predictors can be naturally interpreted in terms of an odds-ratio gradient, $\frac{p}{1-p}$ (e.g.\ the relationship between the odds-ratio of AGN activity and distance to the cluster centre).  
As for our analysis  we employed  an  important extension of the  GLM methodology  known as generalized linear mixed models \citep[GLMMs;][]{hilbe2009logistic}.  The GLMM model, in our context, accounts for  variations in the  intercept and slopes $\beta_{ij}$:
\begin{equation}
\eta_{ij} = \beta_{1,j}+\beta_{2,j}x_1+\cdots+\beta_{kj}x_k,
\label{eq:intercept}
\end{equation}
where the index  $j$  represents the different galaxy morphologies. There are a few key advantages in this more general methodology when compared to the standard approaches. Classical estimation, which separates the information of each group in sub-samples, can be  useless if the sample size in a given  group is too small. On the other hand, if a classical regression is applied ignoring group indicators, the results can be misleading by ignoring group-level variation. GLMMs, also known as  multilevel modelling, represent a compromise  between the overly noisy analysis of each group independently and the oversimplified approach that ignores group indicators \citep{Gelman2007}. A foremost reason to prefer a GLMM model over a standard logistic model relates to the lack of independence between observations when galaxy morphologies are considered as groups. A basic logistic model, whether estimated using maximum likelihood or by using Bayesian methods assumes the independence between each observation in the model. When the data are grouped into levels based on galaxy morphology, this adds extra correlation into the model. A GLMM adjusts for the within-morphology correlation, and therefore is the suitable model for this data.  

The analysis of this work  is  performed using the  Just Another Gibbs Sampler
(\textsc{jags})\footnote{\url{http://CRAN.R-project.org/package=rjags}} package, a
program for analysis of  Bayesian  models using a Markov Chain Monte Carlo
(MCMC) framework. We initiate three  Markov chains by starting the Gibbs
sampler at different initial values sampled from a  Normal distribution with
zero mean and standard deviation of 100. The initial  burn-in phases were set
to $20,000$ steps followed subsequently by $50,000$ iteration steps, which are
sufficient to guarantee the  convergence of each chain via the so-called
Gelman-Rubin statistics \citep{gelman1992}.

\section{Dependence of  Seyfert activity with cluster properties}
\label{sec:res}

To model the presence or absence of Seyfert galaxies, $f_{\rm Seyfert}$, we apply a Bayesian logistic regression. The predictor variables reflecting the environment where the galaxy resides and the host cluster property are  $r/r_{200}$, and $\log M_{200}$\footnote{We standardized the predictors before the  analysis in order to improve possible collinearity and scaling bias due to units differences.}. The $r/r_{200}$ works as a proxy for the local density measure, but can be also understood as a time-scale, given its relationship to the time since the galaxy infall into the cluster began \citep{Gao2004} (modulo a small proportion of `backsplash' galaxies; \citealt{Pim11}). This time-scale is also connected with processes that quench star formation in galaxies. For instance, long-time scale interaction processes, such as strangulation or harassment, would be effective over the entire radial range, while shorter time scale processes, such as ram pressure, would be more localized near the centre, where the density of the gas is higher \citep{vonderLinden.etal:2010}.    

\begin{figure}
\centering
 \resizebox{1.25\hsize}{!}{\includegraphics[trim={1.25cm 1.5cm 0.2cm 1cm},clip]{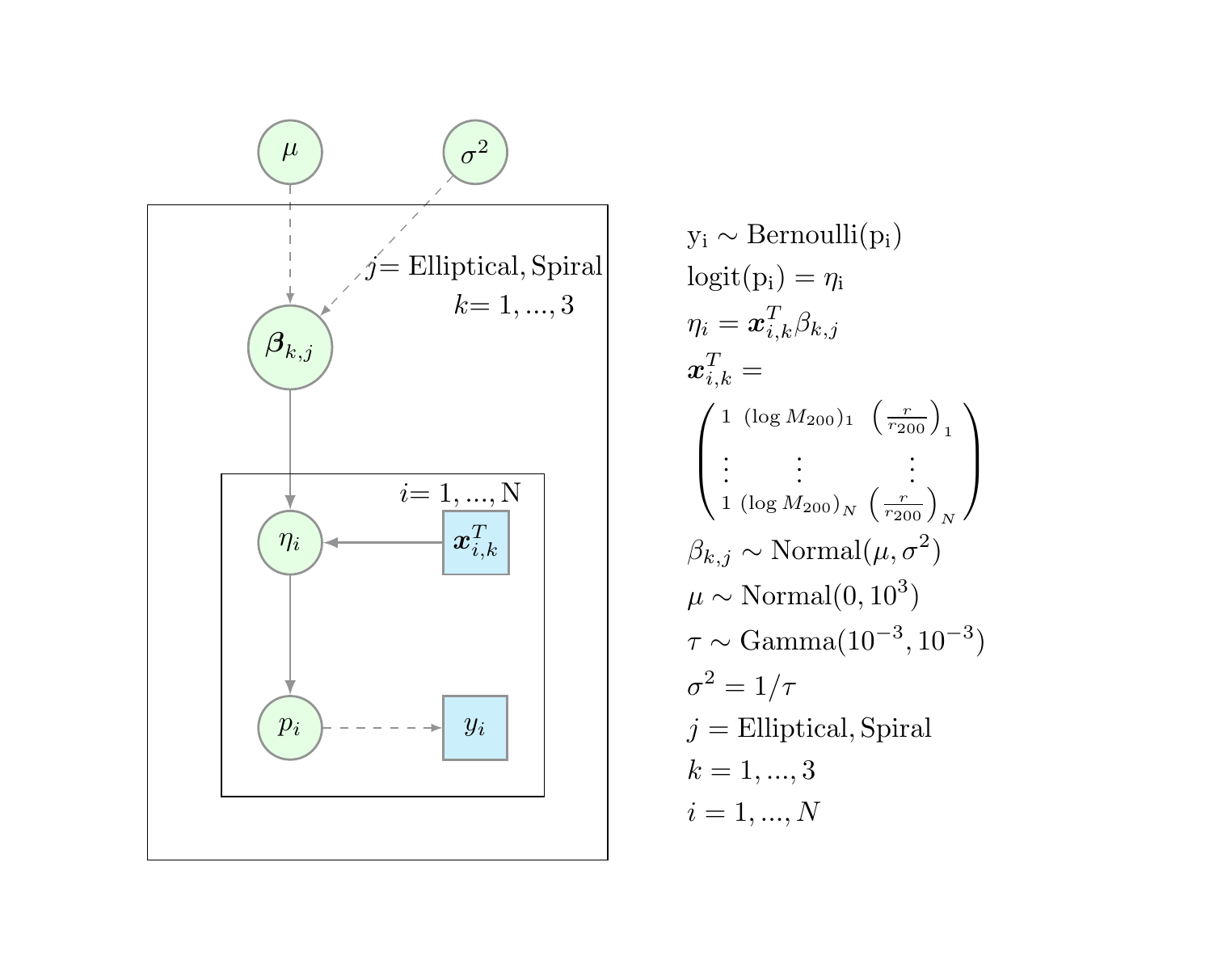}}
\caption{A graphical model representing the hierarchy of  dependencies for a data set of galaxies indexed  by the subscript $i$. The dashed arrows represent stochastic dependencies, while straight arrows the systematic  ones. Blue square represents input data, and green circles are model parameters.}
\label{fig:HBM}
\end{figure}

Our model simultaneously accounts for the dependence of $f_{\rm Seyfert}$ on $r/r_{200}$,  $\log M_{200}$, and galaxy morphology. 
The model is portrayed as a graphical model in  Fig.~\ref{fig:HBM}, and reads as follows: each galaxy in the dataset, composed of  $N$ objects, has its probability of hosting a Seyfert-AGN described by a Bernoulli distribution whose probability of success, $p \equiv f_{\rm Seyfert}$, relates to $r/r_{200}$,  and   $\log M_{200}$ through a logit link function (to ensure the  probabilities will fall  between 0 and 1) and the  linear predictor 
\begin{equation}
\eta = \beta_{1,j} + \beta_{2,j}\cdot \log M_{200} + \beta_{3,j}\cdot r/r_{200},  
\end{equation}
where $j$ is an index representing if a galaxy is elliptical or spiral. 
We assume non-informative priors for the coefficients $\beta_1$, $\beta_2$, $\beta_3$, where we assumed  Normal priors with  mean $\mu$ and standard deviation $\sigma$ for which we assign shared hyper-priors $\mu \sim \mathrm{Normal}(0,10^3)$ and $1/\sigma^2 \sim \mathrm{Gamma}(10^{-3},10^{-3})$.\footnote{The inverse Gamma prior accounts for the fact that  the variance is always positive.} By employing a hierarchical Bayesian model for the varying coefficients $\boldsymbol{\beta_j}$, we allow the model to borrow strength across galaxy types. This happens via their joint influence on the posterior estimates of the unknown hyper-parameters $\mu$ and $\sigma^2$. Thus, the  mixed model herein employed can be understood as a compromise between an analysis that does not account for the information regarding different galaxy morphologies \citep[e.g., Fig. 7;][]{Pimbblet2013}, and the one which splits the data into  independent slices \citep[e.g. Fig 12;][]{vonderLinden.etal:2010}. The former  implicitly assumes a pooled estimate \citep[e.g.,][]{Gelman2007}, i.e. both galaxy types are sampled from the same common distribution ignoring any possible variation among them. The later represents an independent analysis for each class, making the assumption that each morphological type is sampled from independent distributions.  Our GLMM, on the other hand, allows us to account for the differences between elliptical and spiral hosts in an integrated fashion.

Furthermore, by modelling the data on its natural scale (i.e. as a binary variable), our model does not require any arbitrary data  binning,  and our predicted  fractions of AGN are always physically meaningful, even if extrapolated. Such features cannot  be achieved  by standard linear fitting methods. We refer the reader to appendix~\ref{app:JAGS} for the full script explaining how to implement the model in  {\sc jags}. 

\begin{figure*}
\resizebox{0.32\hsize}{!}{\includegraphics{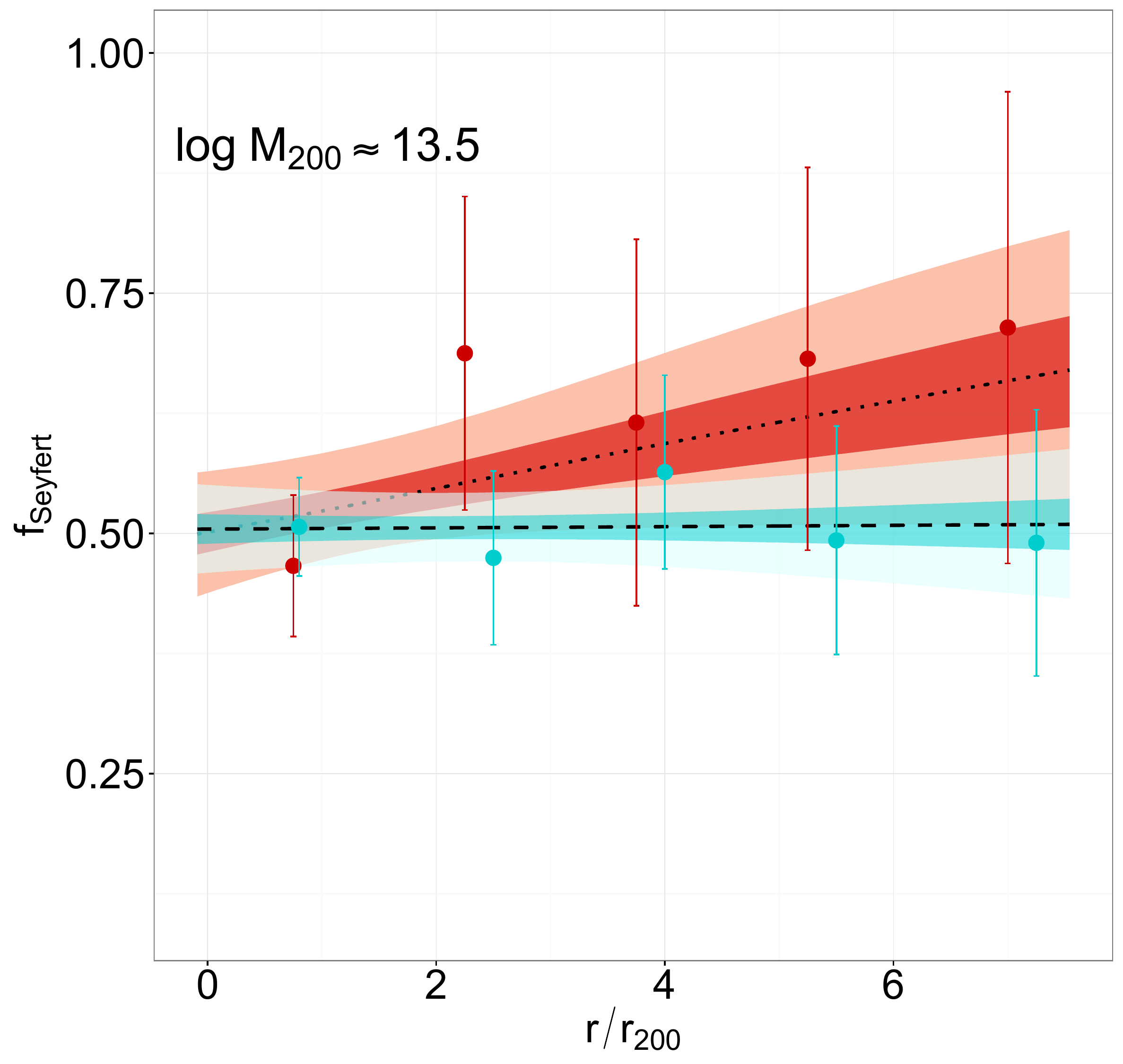}} 
\resizebox{0.32\hsize}{!}{\includegraphics{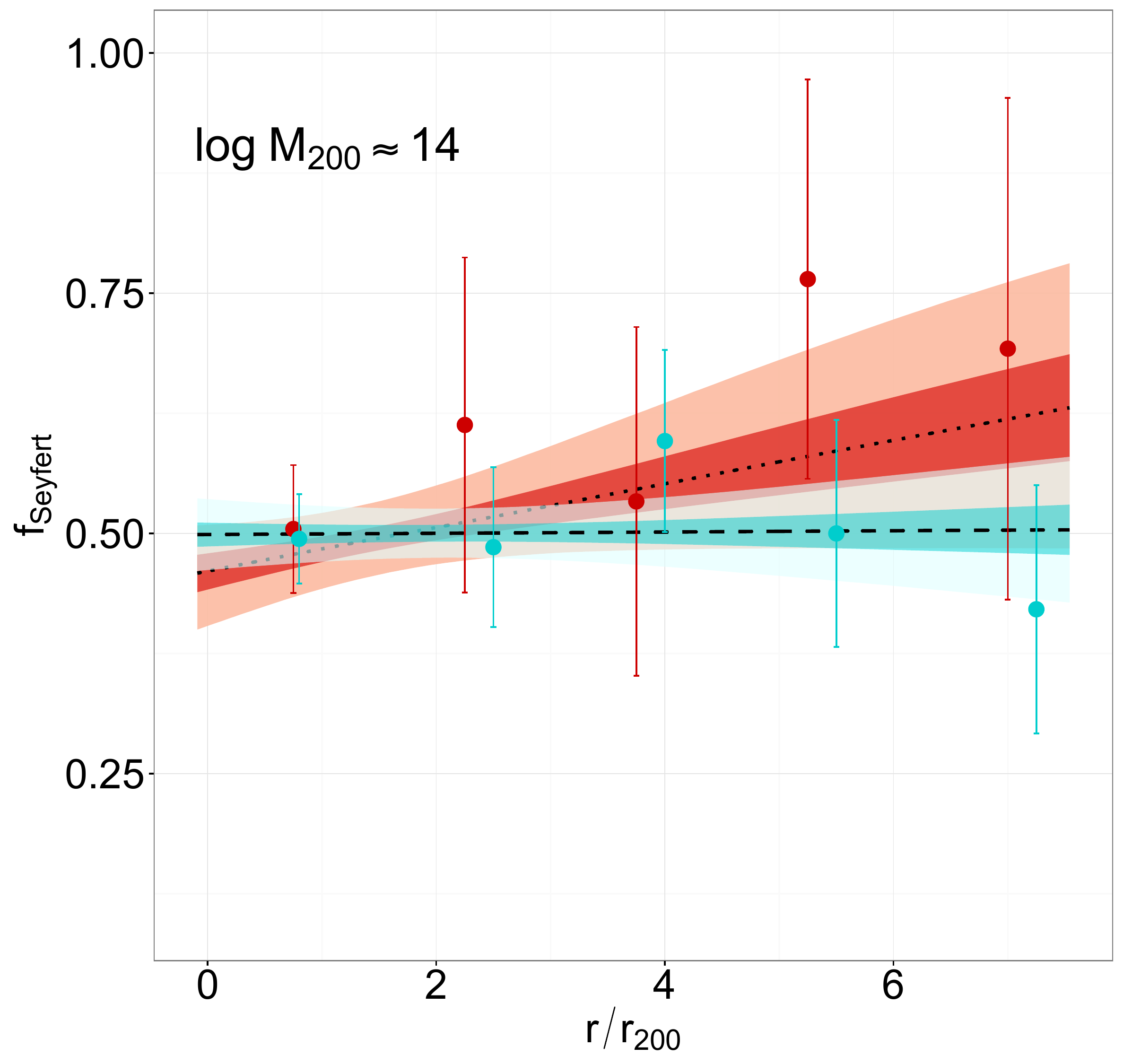}} 
\resizebox{0.32\hsize}{!}{\includegraphics{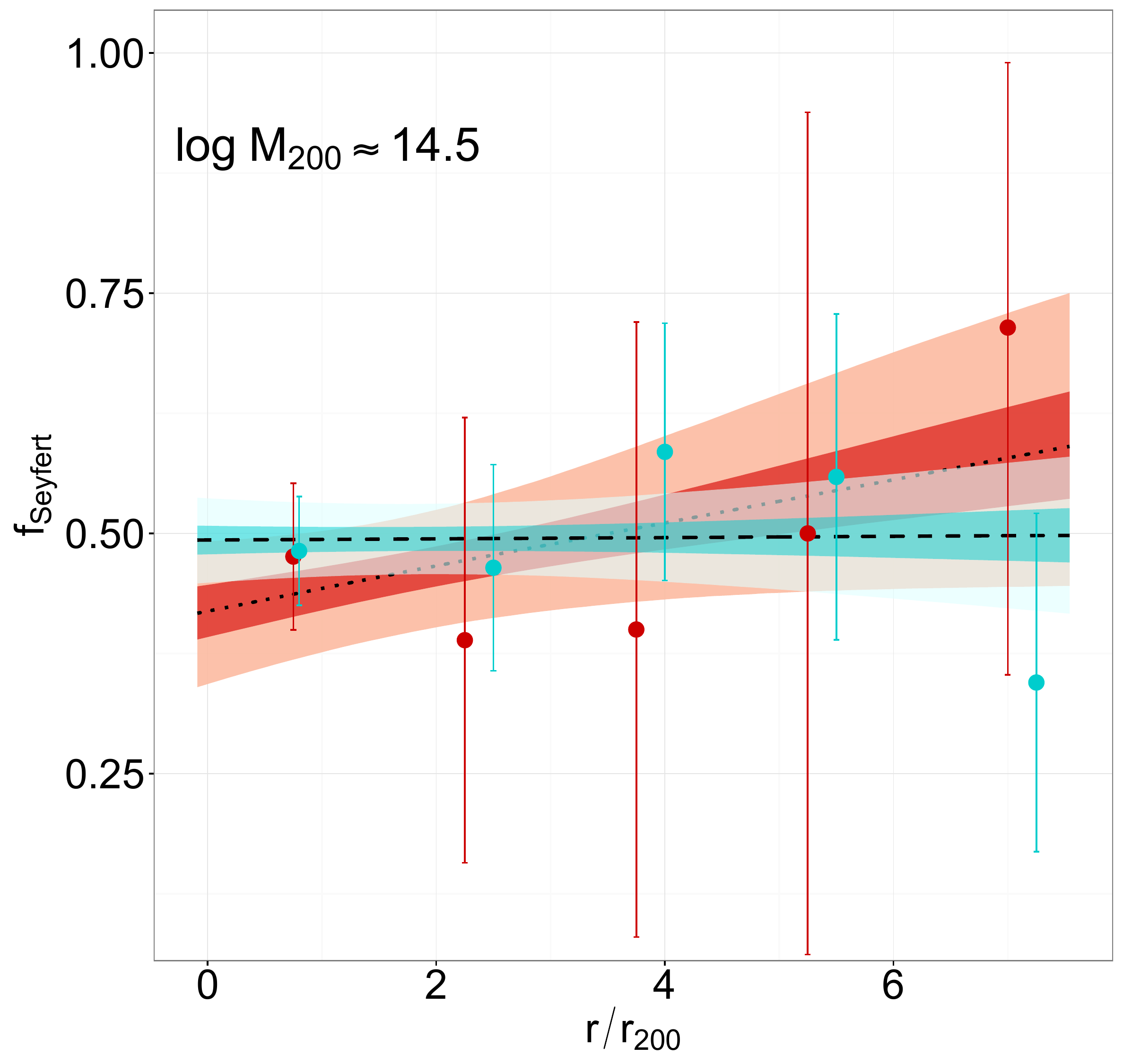}}
\caption{Two-dimensional representation, in slices of $\log M_{200}$,  of the six-dimensional parameter space describing the dependence of Seyfert-AGN activity  as a function of $r/r_{200}$, and $\log M_{200}$  for spirals, and elliptical galaxies. In each panel the black lines  represents the posterior  mean  probability  of Seyfert-AGN activity for each value of $r/r_{200}$, while the shaded areas depicts 50 per cent, and  95 per cent probability intervals. Galaxy types are colour-shape-coded as follows: Ellipticals (red shaded area, dotted line),  spirals (green  shaded area, dashed line). Data points with error bars represent the binned  data (slightly shifted along the $x$-axis for clarity) for purely illustrative purposes.}
\label{fig:logit_fit}
\end{figure*}

The fitted coefficients are displayed in  Tab.~\ref{table:fit}; in
Fig~\ref{fig:parameter_value}, we present their posteriors. To visualise how the model fits the data, we display, in Fig.~\ref{fig:logit_fit}, the predicted probabilities  $f_\mathrm{Seyfert}$ as a function of $r/r_{200}$  in slices of $\log M_{200}$. For each slice, we present the stacked data, for illustrative purposes, and the fitted model and uncertainty. The shaded areas represent 50 per cent, and 95 per cent probability intervals. We recall that the fitting was performed without making use of any data binning. The simple linear model, in the linear predictor scale, is flexible enough to fit the data well, without the need of non-linear dependencies.

The coefficients for the logit model represent the log of the odds ratio for Seyfert activity. Since the predictors are scaled, it allows us to perform a relative comparison between variables measured in different units. For example, 1-$\sigma$  variation in the  $r/r_{200}$ ($\approx 2$) towards the cluster outskirts for a elliptical galaxy residing in a cluster with an average mass $\log M_{200} = 14$ produces on average a change of 0.197 in the log of odds ratio, or in other words it  is 21.7 per cent more likely to be a Seyfert galaxy. Likewise, an elliptical galaxy at an average $r/r_{200}$ ( $\approx 2.2$) residing in a cluster with $\log M_{200} = 14.5$
is $\approx 15.5$ percent less likely to be a Seyfert galaxy than a similar galaxy residing in a cluster with $\log M_{200} = 14$. 
Unlike  elliptical galaxies, spirals are virtually unaffected by the position inside the cluster or  the mass of its host,  with all the  fitted coefficients being consistent with zero.

\begin{table}

\caption{Estimated $\beta_i$ coefficients from the Bayesian logistic regression analysis for ellipticals and spirals galaxies. The $\Delta$Odds  represents the expected change in the odds of Seyfert activity  by a variation of 1-$\sigma$  in the predictor variables, while holding the another at their mean. }
\begin{tabular}{lrcrc}
\hline
\hline 
  & \multicolumn{4}{c}{Galaxy Morphological Type}\\
  & \multicolumn{2}{c}{Ellipticals} & \multicolumn{2}{c}{Spirals} \\
  \cmidrule(lr){2-5}
  &  \multicolumn{1}{c}{$\beta_i$} & $\Delta$Odds  & \multicolumn{1}{c}{$\beta_i$} &  $\Delta$Odds\\
  \hline
 Intercept &   $0.049 \pm 0.088$ &  & $0.003 \pm 0.052$ &  \\
 $  \log M_{200}$ & $-0.169 \pm 0.094$ & -15.5\% & $-0.024 \pm 0.052$ & -2.4\% \\
 $r/r_{200}$ & $0.197 \pm 0.115$ & 21.7\% & $0.006 \pm 0.054$ & 0.6\% \\
  \bottomrule[1.25pt]
\end{tabular}
\label{table:fit}
\end{table}

\section{Discussion and final remarks}
\label{sec:results_discussion}

We study how Seyfert-AGN activity in elliptical and spiral galaxies depends on local and global cluster properties, herein represented by the cluster-centric distance and the mass of the galaxy cluster. For such, we used a sample of low redshift clusters ($0.015 < z < 0.1$), with at least 10 galaxy members each, and with $13.4 < \log M_{200} < 14.6$, from which we include galaxies up to 10$\cdot r_{200}$. It is worth to stress that previous studies have mostly considered only very massive clusters, while  the present work extends the analysis to the lower-end of the cluster mass function.

We found that $f_{\rm Seyfert}$ in elliptical galaxies strongly depends on the $r/r_{200}$, whereas no dependence is seen to occur for the spiral hosts. The deficit of AGN towards the cluster centre is in agreement with previous investigations \citep[e.g.][]{Gilmour2007,Pimbblet2013}, and might be explained by the fact that galaxies in cluster centres should be more stripped of cold gas that can serve as a fuel for AGN activity \citep{Lietzen2011,Pimbblet2013}. Previous studies have shown that conditions in the central regions of clusters are inhospitable for AGN activity even for  galaxies in pairs interactions. The proximity to the cluster centre may   induce decreases in AGN activity by limiting gas availability, most likely of cold gas, around the galaxy \citep{Khabiboulline2014}. This may be caused by the ram pressure stripping \citep{Fujita2004,Roediger2005,Roediger2015}, which for ellipticals may be considerably important, sweeping out the residual gas content of these galaxies, hence starving the central SMBH as consequence \citep{Schawinski.etal:2007,Booth2013,Li2015}. This is also consistent with the  recent finding of \citet{Boselli2014}, that using a volume-limited sample of nearby objects found evidence that ram pressure truncates the infall of pristine gas from the halo, being the dominant process driving the evolution of galaxy clusters.

Nonetheless, a puzzle still remains: why spirals are not effected in the same way?
We argue that the reason might be a consequence of the extra gravitational force provided by their bulges, which dominates the holding force in the central few kpc \citep[see e.g. the \textit{N-body} simulations from][]{Abadi1999}. This is also consistent with the  more recent results from a   hydrodynamical simulation, which includes both mechanical and radiative AGN feedback, from \citet{Shin2012}, which shows that the AGN activity is less affected by gas stripping than the SFR,  because the SMBH accretion is primarily dominated by the density of the galaxy central region.  This phenomenon may be related to the observational  findings of e.g.  \cite{Schawinski2010},  suggesting different characteristics for the evolution of SMBH within spirals and ellipticals. This might be  linked to the positive correlation between the mass of the bulge and the mass of the central SMBH in spirals galaxies \citep{FerrareseMerritt2000,HaringRix2004,Somerville2008,Reines2015}, which provides the extra potential well in the few central kpc. Thus, the spirals might be capable of holding the internal reservoir of gas independently of environment effects, thus keeping the fraction of nuclear activity nearly constant throughout their position inside the cluster.

Our analysis and results can be summarized as follows:
\begin{itemize}

\item  To control for effects of confounding variables, we built a control sample of inactive galaxies matching each pair against their $M_{*}$, $(g-r)$, and SFR, hence allowing us to make cleaner claims about causal effects;

\item The Bayesian logistic model does not need to rely on arbitrary data binning, providing the proper scale for modelling a binary variable (presence/absence of Seyfert). Therefore allowing an intuitive interpretation of the fitted coefficients, and unbiased results when compared to standard Gaussian approaches;   

\item The odds of Seyfert-AGN in elliptical hosts increases towards the group/cluster outskirts  by a factor of $\sim 21$ per cent  for one-$\sigma$  increase of $r/r_{200}$, in clusters with an average mass of $\log M_{200} \approx 14$, and the effect is more prominent in more massive clusters; 

\item The  mixed model used in this work accounts for the differences between elliptical and spiral galaxies into  an integrated fashion, borrowing  strength across sub-samples; 

\item The $f_{\rm Seyfert}$ in spirals are virtually unaffected by the external environment. We argue that this can be a consequence of the spirals being able to hold an internal reservoir of gas, and this might be caused by some type of protection provided by the galaxy's disc or by extra gravitational force provided by their central regions.

\item An alternative scenario for our findings  may represent a challenge for the unified  AGN model. The different behaviour of AGN in spiral and elliptical hosts  may  indicate that not all AGN are  the born same, and may depend of their surrounding environment.  This is consistent with \cite{Villarroel2014}, who found differences in the host morphologies behaviour of Type-1 and Type-2 AGN and indicatives for two distinct classes. An interesting puzzle that shall be addressed in a follow-up investigation.

\end{itemize}

Generalized linear models are a cornerstone of modern statistics, but nearly {\it Terra incognita} in astronomical investigations.    In the present work, we employed a Bayesian logistic  mixed  model  designed to represent binary data, to analyse the probability of Seyfert-AGN activity as function of their position inside the group/cluster and the mass of its host. Standard approaches in the literature applying a classical linear regression in a fractional data (e.g. fractions of AGN per bin of $r/r_{200}$) do not have any restriction requiring the prediction fractions to fall between 0 and 1,  and cannot be extrapolated while still  providing meaningful results. The combination of a propensity score matching and a Bayesian logistic mixed regression allow us to model  the $f_{\rm Seyfert}$ with an unparalleled statistical robustness up to date.

Finally we are able to answer the question posed in the title: \textit{Is the cluster environment quenching the Seyfert  activity in elliptical and spiral galaxies?} In elliptical galaxies, yes. The regions towards the cluster centre are hostile to the central SMBH residing in theses galaxies. On the other hand, spirals appear to be spared and they seem to keep feeding their AGN while falling into the cluster potential.

\section*{Acknowledgements}

This work is a product of the 2$^{\rm nd}$ COIN Residence Program. We thank
Alan Heavens and Jason McEwen for encouraging the realization of this edition. We thank E. E. O. Ishida, L. Dobos, R. Beck,  B. Kocsis, and B. Villarroel for the fruitful discussions and suggestions. A particular thank goes for M. Trevisan for kindly providing the dataset used in this work and for participating in several discussions. We thank the anonymous referee for the constructive suggestions and comments.
The program was held in the Isle of Wight, UK in October/2015 and supported by
the Imperial Centre for Inference and Cosmology (ICIC), Imperial College of
London, UK, and by the Mullard Space Science Laboratory (MSSL) at the
University College of London, UK.
The IAA Cosmostatistics
Initiative\footnote{\url{https://asaip.psu.edu/organizations/iaa/iaa-working-group-of-cosmostatistics}}
(COIN) is a non-profit organization whose aim is to nourish the synergy between
astrophysics, cosmology, statistics, and machine learning communities. This work
was written on the collaborative {\sc overleaf} 
platform\footnote{\url{www.overleaf.com}}, made use of
{\sc github}\footnote{\url{https://github.com}} a web-based hosting service and
\texttt{git} version control software,
{\sc datajoy}\footnote{\url{www.getdatajoy.com}} an on-line collaborative programming
platform for {\sc  python} and {\sc r} users,
and {\sc slack}\footnote{\url{https://slack.com}} a team collaboration platform.



\appendix

\section{Posteriors}
\begin{figure}
\resizebox{0.95\hsize}{!}{\includegraphics{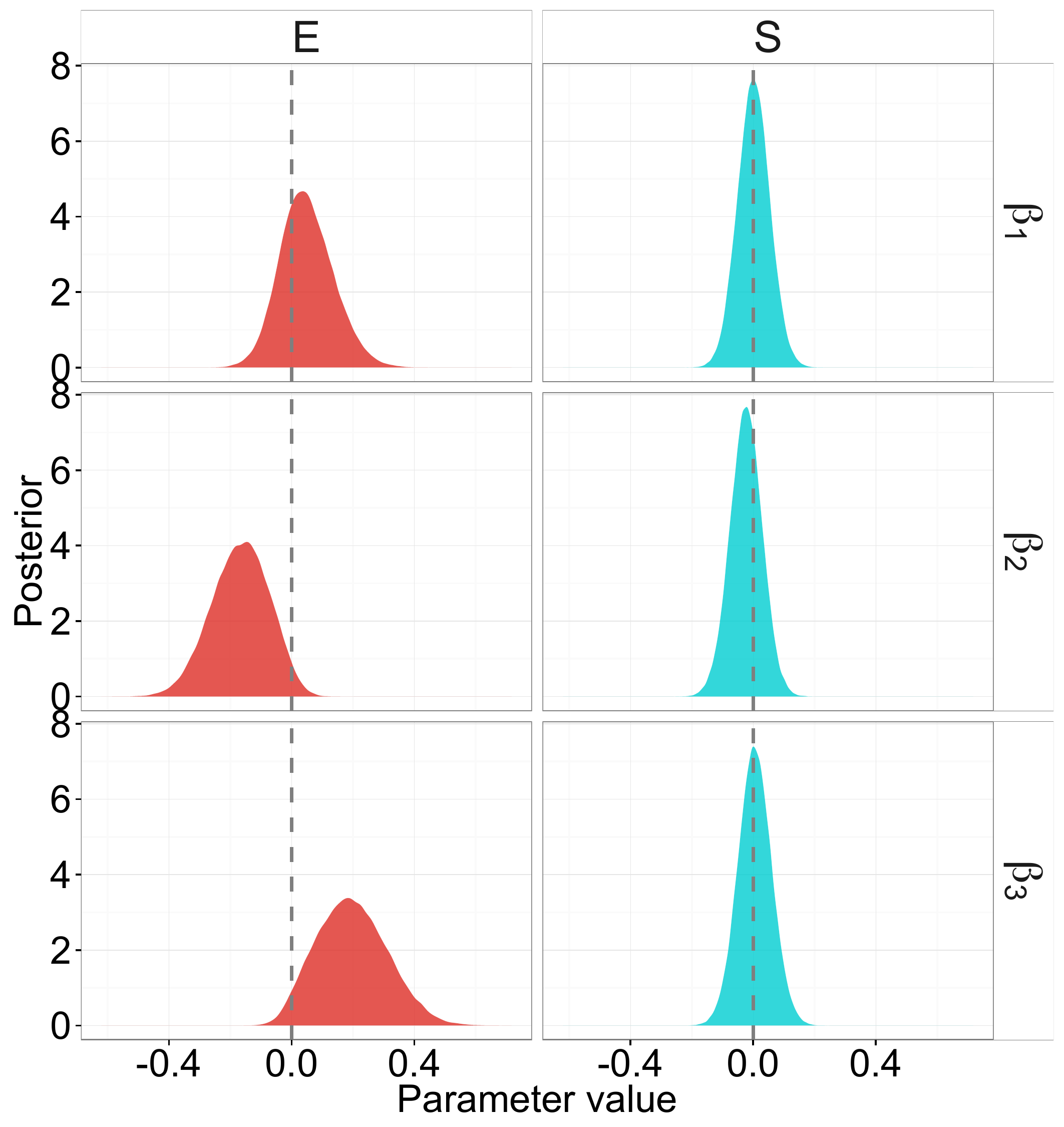}} 
\caption{Computed posterior for the $\beta$ coefficients of Fig.~\ref{fig:HBM}
From top to bottom $\beta$ stands for: Intercept, $\log M_{200}$, and
$r/r_{200}$ for elliptical (left panel) and spiral (right panel) galaxies
respectively.}
\label{fig:parameter_value}
\end{figure}

We display in Figure~\ref{fig:parameter_value} the computed posteriors for the $\beta$ coefficients in Equation  confidence interval.

\section{JAGS model}
\label{app:JAGS}
\lstset{
language=R,
keywordstyle= \color{black},
identifierstyle=\ttfamily,
commentstyle=\color{orange!30!red},
stringstyle=\ttfamily\color{blue!75!green},
xleftmargin=1pt,
framexleftmargin=20pt,
showstringspaces=false,
basicstyle=\footnotesize,
numbersep=10pt,
tabsize=2,
breaklines=true,
prebreak = \raisebox{0ex}[0ex][0ex]{\ensuremath{\hookleftarrow}},
breakatwhitespace=false,
aboveskip={1.5\baselineskip},
columns=flexible,
extendedchars=true
}

In the following, we show how to translate the graphical model in
Fig.~\ref{fig:HBM}  into {\sc jags} using the {\sc r2jags} package. Note that
{\sc jags} uses precision $\tau = 1/\sigma^2$ in place of variance $\sigma^2$.
The whole script,  which one should run from within {\sc r},  to load the
data  and recover the coefficients displayed in Tab.~\ref{table:fit}  is given
below:

\begin{lstlisting}
#######################################
# JAGS Script in R
# Required libraries
library(R2jags)
#Data
data<-read.csv("https://goo.gl/ppMoSl",header=T)

X <- model.matrix( ~  logM200 + r_r200, data = data) 
K <- ncol(X)                #Number of Predictors
y <- data$bpt               #Response variable 
n <- length(y)              #Sample size
gal <- as.numeric(data$zoo) #Galaxy type

#JAGS data
jags_data  <- list(Y= y,
                   N = n,
                   X=X,
                   gal=gal)
#Model
jags_model<-"model{
#Shared hyperpriors for beta 
tau ~ dgamma(1e-3,1e-3)     #Precision
mu  ~ dnorm(0,1e-3)         #mean
#Diffuse prior for beta 
for(j in 1:2){
for(k in 1:3){
beta[k,j]~dnorm(mu,tau)                                    
}}
# Likelihood
for(i in 1:N){
Y[i] ~ dbern(pi[i])
logit(pi[i]) <-  eta[i]
eta[i] <- beta[1,gal[i]]*X[i,1]+
beta[2,gal[i]]*X[i,2]+
beta[3,gal[i]]*X[i,3]
}}"
# Monitor this parameter     
params <- c("beta") 
# Generate initial values
inits  <- function () {
list(beta = matrix(rnorm(6,0, 0.01),ncol=2))} 
# Run MCMC
burn    = 2*10^4            #Burn-in samples
s       = 5*10^4            #Number of samples
nc      = 3                 #Number of mcmc
th      = 10                #Thinning value

jags_fit <- jags(
data = jags_data,
inits      = inits,
parameters = params,
model.file = textConnection(jags_model),
n.chains   = nc,
n.thin     = th,
n.iter     = s,
n.burnin   = burn)

#Partial output
print(jags_fit,intervals=c(0.025, 0.975), 
digits=3)
\end{lstlisting}
An example of the output of the \textsc{jags} model
is shown below.
\begin{lstlisting}[language=bash]
#######################################
Inference for Bugs model, fit using jags,
3 chains, each with 50000 iterations
           mu.vect sd.vect     2.5%    97.5%  
beta[1,1]    0.049   0.088   -0.112    0.238 
beta[2,1]   -0.169   0.094   -0.356    0.003 
beta[3,1]    0.197   0.115   -0.010    0.428 
beta[1,2]    0.003   0.052   -0.100    0.107 
beta[2,2]   -0.024   0.052   -0.130    0.075 
beta[3,2]    0.006   0.054   -0.099    0.113
#######################################
\end{lstlisting} 
\bsp 
\label{lastpage}
\end{document}